
\newfam\scrfam
\batchmode\font\tenscr=rsfs10 \errorstopmode
\ifx\tenscr\nullfont
	\message{rsfs script font not available. Replacing with calligraphic.}
\else	\font\sevenscr=rsfs7
	\font\fivescr=rsfs5
	\skewchar\tenscr='177 \skewchar\sevenscr='177 \skewchar\fivescr='177
	\textfont\scrfam=\tenscr \scriptfont\scrfam=\sevenscr
	\scriptscriptfont\scrfam=\fivescr
	\def\scr{\fam\scrfam}
	\def\cal{\scr}
\fi
\newfam\msbfam
\batchmode\font\twelvemsb=msbm10 scaled\magstep1 \errorstopmode
\ifx\twelvemsb\nullfont\def\Bbb{\bf}
	\message{Blackboard bold not available. Replacing with boldface.}
\else	\catcode`\@=11
	\font\tenmsb=msbm10 \font\sevenmsb=msbm7 \font\fivemsb=msbm5
	\textfont\msbfam=\tenmsb
	\scriptfont\msbfam=\sevenmsb \scriptscriptfont\msbfam=\fivemsb
	\def\Bbb{\relax\ifmmode\expandafter\Bbb@\else
 		\expandafter\nonmatherr@\expandafter\Bbb\fi}
	\def\Bbb@#1{{\Bbb@@{#1}}}
	\def\Bbb@@#1{\fam\msbfam\relax#1}
	\catcode`\@=\active
\fi
%
%
\font\eightrm=cmr8	\def\xrm{\eightrm}
\font\eightbf=cmbx8	\def\xbf{\eightbf}
\font\eightit=cmti8	\def\xit{\eightit}
\font\eighttt=cmtt8	\def\xtt{\eighttt}
\font\eightcp=cmcsc8
\font\tencp=cmcsc10
\font\tentt=cmtt10
\font\twelverm=cmr12
\font\twelvecp=cmcsc10 scaled\magstep1
\font\fourteencp=cmcsc10 scaled\magstep2
\def\ss{\scriptstyle}
%
%
\headline={\ifnum\pageno=1\hfill\else\hfill
{\eightcp Cederwall, von Gussich, Sundell:
	Deformations in Closed String Theory}\dotfill\folio\fi}
\def\makeheadline{\vbox to 0pt{\vss\noindent\the\headline\break
\hbox to\hsize{\hfill}}
	\vskip2\baselineskip}
%
%
\footline={}
\def\makefootline{\baselineskip=1.6cm\line{\the\footline}}
%
%
\newcount\sectioncount
\sectioncount=0
\def\section#1{\global\eqcount=0
	\global\advance\sectioncount by 1
	\vskip2\baselineskip\noindent
	\hbox{\twelvecp\the\sectioncount. #1\hfill}\vskip\baselineskip}
\def\appendix#1{\vskip2\baselineskip\noindent
	\hbox{\twelvecp Appendix: #1\hfill}\vskip\baselineskip}
%
%
\newcount\refcount
\refcount=0
\newwrite\refwrite
\def\ref#1#2{\global\advance\refcount by 1
	\xdef#1{\the\refcount}
	\ifnum\the\refcount=1
	\immediate\openout\refwrite=\jobname.refs
	\fi
	\immediate\write\refwrite
		{\item{[\the\refcount]} #2\hfill\par\vskip-2pt}}
\def\refout{\catcode`\@=11
	\immediate\closeout\refwrite
	\vskip2\baselineskip
	{\noindent\twelvecp References}\hfill\vskip\baselineskip
	\parskip=.875\parskip
	\baselineskip=.8\baselineskip
	\input\jobname.refs
	\parskip=8\parskip \divide\parskip by 7
	\baselineskip=1.25\baselineskip 
	\catcode`\@=\active}
%
%
\newcount\eqcount
\eqcount=0
\def\Eqn#1{\global\advance\eqcount by 1
	\xdef#1{\the\sectioncount.\the\eqcount}
		\eqno(\the\sectioncount.\the\eqcount)}
\def\eqn{\global\advance\eqcount by 1
	\eqno(\the\sectioncount.\the\eqcount)}
%
%
\parskip=3.5pt plus .3pt minus .3pt
\baselineskip=12pt plus .1pt minus .05pt
\lineskip=.5pt plus .05pt minus .05pt
\lineskiplimit=.5pt
\abovedisplayskip=8pt plus 4pt minus 3.4pt
\belowdisplayskip=\abovedisplayskip
\hsize=15cm
\vsize=20.2cm
\hoffset=1cm
\voffset=1.1cm
%
%
\def\zbar{\bar z}
\def\wbar{\bar w}
\def\/{\over}
\def\d{\partial}

\def\a{\alpha}
\def\b{\beta}
\def\e{\varepsilon}
\def\g{\gamma}
\def\r{\varrho}
\def\s{\sigma}
\def\G{\Gamma}
\def\A{\hbox{A}}
\def\Z{{\Bbb Z}}
\def\C{{\Bbb C}}
\def\R{{\cal R}}
\def\V{{\cal V}}
\def\W{{\cal W}}
\def\rpr{\r_\Phi(R)}
\def\rqr{\r_\Psi(R)}
\def\dpr{\delta_\Phi(R)}
\def\ad{\hbox{ad}}
\def\punkt{\,\,.}
\def\komma{\,\,,}
\def\.{.\hskip-1pt }
\def\is{\!=\!}
\def\-{\!-\!}
\def\+{\!+\!}
\def\Int{\int\limits}
\def\bra{\,<\!\!}
\def\ket{\!\!>\,}
\def\F{\,{}_2F_1}
\def\FF{\,{}_3F_2}

\def\Res{\hbox{Res}}
\def\oop{{\lower2pt\hbox{$1$}\/\raise2pt\hbox{$\pi$}}\!}
\def\oopsq{{\lower2pt\hbox{$1$}\/\raise2pt\hbox{$\pi^2$}}\!}
\def\Re{\hbox{Re}\,}
\def\Im{\hbox{Im}\,}
%
%
%
%
%
\null\vskip-1cm
\hbox to\hsize{\hfill G\"oteborg-ITP-95-8}
\hbox to\hsize{\hfill\tt hep-th/9504112}
\hbox to\hsize{\hfill April 1995}
\hbox to\hsize{\hfill revised, December 1995}

\vskip3cm
\centerline{\fourteencp Deformations in Closed String Theory ---}
\vskip\parskip
\centerline{\fourteencp Canonical Formulation and Regularization}

\vskip1.5cm
\centerline{\twelverm Martin Cederwall, Alexander von Gussich
	and Per Sundell}

\vskip1.5cm
\centerline{\it Institute for Theoretical Physics}
\centerline{\it Chalmers University of Technology and G\"oteborg University}
\centerline{\it S-412 96 G\"oteborg, Sweden}

\vskip3.5cm
\noindent \underbar{Abstract:} We study deformations of closed string theory
by primary fields of conformal weight $(1,1)$,
using conformal techniques on the complex plane. A canonical surface integral
formalism for computing commutators in a non-holomorphic theory is constructed,
and explicit formul\ae\ for deformations of operators are given.
We identify the unique regularization of the arising divergences
that respects conformal invariance, and consider the corresponding parallel
transport. The
associated connection is metric compatible and
carries no curvature.

\vskip2.5cm
\vfill
\catcode`\@=11
\hbox to\hsize{email addresses: \tentt tfemc@fy.chalmers.se\hfill}
\hbox to\hsize{\phantom{email addresses: }\tentt tfeavg@fy.chalmers.se\hfill}
\hbox to\hsize{\phantom{email addresses: }\tentt tfepsu@fy.chalmers.se\hfill}
\catcode`\@=\active

\eject

\ref\Zwiebach{B.~Zwiebach, \xit Nucl.Phys\.\ \xbf B390
		\xrm (1993) 33, {\xtt hep-th/9206084};\vskip-3pt
	\item{}B.~Zwiebach, {\xtt hep-th/9305026}.}
\ref\vongussichsundell{A.~von Gussich and P.~Sundell, in preparation.}
\ref\Rang{K.~Ranganathan, \xit Nucl.Phys\.\ \xbf B408 \xrm (1993) 180,
	{\xtt hep-th/9210090}.}
\ref\RSZ{K.~Ranganathan, H.~Sonoda and B.~Zwiebach,
	\xit Nucl.Phys\.\ \xbf B414 \xrm (1994) 405, {\xtt hep-th/9304053}.}
\ref\Pelts{G.~Pelts, \xit Phys.Rev. \xbf D51 \xrm (1995) 671,
	{\xtt hep-th/9406022}.}
\ref\kugozwiebach{T. Kugo, B. Zwiebach, \xit Prog.Theor.Phys\.\ \xbf 87
	\xrm (1992) 801-860.}
\ref\Hawking{S.~Hawking, \xit Commun.Math.Phys\.\ \xbf 55 \xrm (1977) 133.}
\ref\EOi{M.~Evans and B.~Ovrut, \xit Phys.Rev\.\ \xbf D39 \xrm (1989) 3016.}
\ref\CNW{M.~Campbell, P.~Nelson and E.~Wong,
	\xit Int.J.Mod.Phys\.\ \xbf A6 \xrm (1991) 4909.}
\ref\EGN{M.~Evans, I.~Giannakis and D.V.~Nanopoulos,
	\xit Phys.Rev. \xbf D50 \xrm (1994) 4022,
	{\xtt hep-th/9401075}.}
\ref\Zuck{G.J.~Zuckerman, talk presented at the
	$\scriptstyle 1^{\scriptscriptstyle\underline{st}}$
	Feza G\"ursey Memorial
	Conference\hfill\break\indent on  Strings and Symmetries, Istanbul,
	Turkey, 1994.}
\ref\EOii{M.~Evans and B.~Ovrut, \xit Phys.Rev\.\ \xbf D41 \xrm (1990) 3149.}
\ref\DDF{E.~Del~Giudice, P.~Di~Vecchia and S.~Fubini,
	\xit Ann.Phys\.\ \xbf 70 \xrm (1972) 378.}
\ref\Heterotic{D.~Gross, J.~Harvey, E.~Martinec and R.~Rohm, \xit Nucl.Phys\.\
	\xbf B267 \xrm (1986) 75.}
\ref\Hyper{H.~Exton, ``Handbook of Hypergeometric Integrals'', Ellis Horwood,
		Chichester, 1978;\vskip-3pt
	\item{} M.~Abramowitz and I.A.~Stegun (eds.),
		``Handbook of Mathematical Functions'',\hfill\break\indent
		National Bureau of Standards, Washington, 1972.}
\ref\Background{A.~Sen and B.~Zwiebach, \xit Nucl.Phys\.\ \xbf B414
		\xrm (1994) 649, {\xtt hep-th/9307088};\vskip-3pt
	\item{}A.~Sen and B.~Zwiebach, \xit Nucl.Phys\.\ \xbf B423
		\xrm (1994) 580, {\xtt hep-th/9311009}.}

\section{Introduction and Summary}
Probably the most important theoretical problem concerning string theory
is the lack of a ``covariant'' formulation. Despite the fact that closed
string theory contains gravity as part of the infinite spectrum, there
is no formulation of string theory that is manifestly invariant under
general coordinate transformations. It is likely that a fundamental
gauge principle of closed string theory involves some quantum geometric
invariance generalizing that of Einstein's theory of gravity.
The most promising place to look for such gauge symmetries is closed
string field theory [\Zwiebach]. However, despite the interesting algebraic
structures arising in the field theory formulation of string theory,
some aspects are still simpler in a first-quantized version. A scattering
amplitude that in the first-quantized framework is given by a single
integral with vertex operator insertions decomposes into several terms
in the field theory, due to the somewhat arbitrary decomposition of a
Riemann surface into propagator and vertex parts.
The present work is performed entirely in a first-quantized formalism,
but it is possible to translate it into a field theoretic language.
We will address that issue in a forthcoming paper [\vongussichsundell],
and only comment on the connection in this paper.

The purpose of the paper is to investigate the local structure of ``string
theory space'', i.e., the space of consistent backgrounds for string
propagation. To this end, we consider ``deformations'' of closed string
theory, where the flat background is shifted to some infinitesimal field
configuration corresponding to physical states in the string theory,
in which a free string propagates.

It is a priori difficult to judge what gauge transformations should look
like, simply because we do not know what the ultimate off-shell
field content is. However, the situation may be better than expected.
We will demonstrate that deformations of closed string theory, i.e.,
transformations between inequivalent backgrounds, are ``almost inner
automorphisms'' of the conformal field theory --- they are generated
by the action of an operator formed from the oscillators in the theory
itself in a given background, and behave analytically on momentum
operators. Each such operator has a regular action on almost all
operators in the theory --- there are only simple poles for
certain momentum eigenvalues (resonances).

In order to construct these operators and investigate their action,
we determine how the apparent divergences arising at various stages
in the calculations should be regularized. The guideline is conformal
invariance, and the answer is unique: analytic continuation. Only
logarithmic divergences survive and produce simple poles, all other
divergences are uniquely regularized to give finite results.
This is an important issue to settle; there has been questions both
about the freedom to choose regularization and what it should look
like [\Rang,\RSZ,\Pelts,\kugozwiebach]. In the previous formulations
the simple poles are regarded as nonregular terms which
need to be subtracted. This introduces an arbitrariness in the
choice of finite counterterms constrained by conformal invariance,
though some preferred connections have been found.
The subtractions are not necessery if $L_{0}$ has
a continous spectrum. Then
the produced simple
poles have a well defined meaning as distributions in the external
momentum decomposition of the background perturbation. This is a
physical assumption which is analogous to the treatment of
the delta functions in the scattering matrix using adiabatic
turn on of the perturbation. In this picture
the perturbation is localized in space-time such that the incoming
and outgoing states are unperturbed.
Regularization by analytic continuation has also
been considered in four dimensions, and has been shown to be equivalent
to dimensional regularization [\Hawking].

Since we will use a canonical framework, we will frequently be calculating
commutators of operators in the non-holomorphic conformal field theory.
To streamline these calculations, we develop a formalism that translate
operator products into commutators, analogously to what is done in
a holomorphic theory with contour integrals (our generators are naturally
surface integrals of local operators). This makes it possible to take
advantage of the simple form of correlation functions for radial quantization
on the complex plane. This property is lost when one goes to equal-time
quantization in cylindrical coordinates, and one is naturally lead to
calculations involving distributions.

Deformations of closed string theory with physical vertex operators as
automorphisms of the double Virasoro algebra were first considered in
[\EOi]. In cylindrical coordinates they take the form
$$
\delta_\Phi T(\s)=\delta_\Phi\bar T(\s)=\Phi(\s)\komma\eqn
$$
where $\Phi$ is a physical vertex operator, a primary field of weight $(1,1)$.
If one goes to planar variables, one obtains
$$\eqalign{
(\delta_\Phi T)(z,\zbar)&={\zbar\/z}\Phi(z,\zbar)\komma\cr
(\delta_\Phi\bar T)(z,\zbar)&={z\/\zbar}\Phi(z,\zbar)\komma\cr}
\eqn
$$
where the factors in front of $\Phi$ arise because of the differences in
conformal weights between $\Phi$ and $T$ or $\bar T$, respectively.
The most ``covariant'' formulation has been given in [\CNW], where
deformations of surface states of given genus and number of punctures
are expressed in terms of surface states of the same genus and one more
puncture. The $(1,1)$-form $\phi\is\Phi(z,\zbar)dz\!\wedge\!d\zbar$ is
inserted at the extra puncture and its position is integrated over.
This formulation is the one that is most suited for string field theory.
Then the $N$-string vertex at genus $g$ gets a modification that comes
from the $(N\+1)$-string vertex at genus $g$. The transformations act
naturally on the second-quantized (multi-string) Fock space, and the problem
with potential divergences from colliding punctures is pushed ahead.
In this paper we work in a first-quantized framework, where the divergences are
taken
care of immediately, without cutting out semi-infinite propagators
from the amplitudes. The transformations act on the first-quantized
(one-string) Hilbert space, and the deformed theory is seen as a single string
moving in a non-trivial background. However, we find relations for amplitudes
that, thanks to the regularization we use, provide a natural link
to a second-quantized formalism. We will adress this in more detail
in the forthcoming paper [\vongussichsundell].

When we consider repeated tranformations, we find that the commutator
of two deformations vanishes identically on the full state space, i.e.
if the deformation is regarded as a parallell transport along
directions in the space of conformal field theories then the curvature
vanishes.

We think that the simple technique we have developed for calculating
commutators in a non-holomorphic theory has a potential to solve
problems associated with closed string theory, especially those connected
to non-holomorphic, ``bilateral'' operators [\EGN,\Zuck]. We intend to
use it in the search for generalizations of the general coordinate
invariance [\EOii], possibly involving operators at all mass levels.
It should also be suited for posing questions about finite deformations
(finite parallell transport)
and more general deformations into non-conformal theories.

\section{Canonical Formulation}
Under the deformation by the primary $(1,\!1)$-field $\Phi(z,\!\zbar)$
the holomorphic and anti-holomorphic components of the stress
tensor transform as
$$
\eqalign{
&T(z)\longrightarrow T(z)+\e{\zbar\/z}\Phi(z,\zbar)\komma\cr
&\bar T(\zbar)\longrightarrow \bar T(\zbar)+\e{z\/\zbar}
	\Phi(z,\zbar)\punkt\cr}\Eqn\Tvar
$$
It is of course interesting to see if these transformations can be
seen as inner derivations (infinitesimal inner automorphisms),
i.e., if they are generated by the adjoint
action of some generator $\r_\Phi$ constructed from the fields in the
theory.

Consider a general physical vertex operator $\Phi(z,\zbar)$, carrying
(left and right) momentum $k$. It can be given a mode expansion as
$$
\Phi(z,\zbar)=\sum_{m,n\in\Z}
	\Phi_{mn}|z|^{2\g}z^{-1-m}\zbar^{-1-n}\komma
	\Eqn\modeexp
$$
where $\g$ is the operator valued shift ${(k\!\cdot\!p)/4}$.
The stress tensor has the expansions
$$
\eqalign{T(z)&=\sum_{m\in\Z}L_mz^{-2-m}\komma\cr
	\bar T(z)&=\sum_{m\in\Z}\bar L_m\zbar^{-2-m}\punkt\cr}\eqn
$$
Since the physical vertex operators factorize in holomorphic and
anti-holomorphic parts, we actually have $\Phi_{mn}=V_m\bar V_n$, but
that expression will not be needed here.
The commutators of the modes of $T$ or $\bar T$ and $\Phi$ are
$$
\eqalign{[L_m,\Phi_{nl}]&=(\g-n)\Phi_{n+m,l}\komma\cr
	[\bar L_m,\Phi_{nl}]&=(\g-l)\Phi_{n,l+m}\punkt\cr
	}\Eqn\modecomm
$$
One may try to construct from the modes of $\Phi$ an operator whose adjoint
action gives the variations (\Tvar) in the stress tensor.
Upon doing this, one must remember that commutators have to be evaluated at
``equal time'', here meaning equal radius. We denote this radius by $R$,
and the generator of deformations corresponding to the field $\Phi$ at
radius $R$ by $\rpr$. We also let $\dpr\is\ad\rpr$.

Expansion of the transformations (\Tvar) on the circle $|z|\is R$ reads
$$
\eqalign{
	\dpr L_m&=\oint\limits_{|z|=R}{dz\/2\pi i}
			z^{1+m}{\zbar\/z}\Phi(z,\zbar)
		=\sum_{n\in\Z}R^{2(\g-n)}\Phi_{n+m,n}\komma\cr
	\dpr\bar L_m&
		=\oint\limits_{|z|=R}{d\zbar\/2\pi i}\zbar^{1+m}
			{z\/\zbar}\Phi(z,\zbar)
		=\sum_{n\in\Z}R^{2(\g-n)}\Phi_{n,n+m}\punkt\cr}\eqn
$$
It is worth stressing that the deformed components of the stress tensor
do not respect any holomorphicity conditions. The actual forms of the deformed
$L_m$ and $\bar L_m$ depend on the radius $R$. This is natural --- their
unitary time evolutions are governed by the deformed time-dependent hamiltonian
$L_0+\bar L_0+\dpr(L_0+\bar L_0)$. It is easily verified that
the new $L_m$'s and $\bar L_m$'s satisfy
$\hbox{\it Vir}\oplus\!\!\hbox{\it Vir}$,
i.e., $\dpr$ is a derivation of the double Virasoro algebra.

Using (\modecomm), we notice that the variations (\Tvar) are formally
generated by
$$
\rpr=-\sum_{n\in\Z}{R^{2(\g-n)}\/\g-n}\Phi_{nn}\punkt\Eqn\sumrho
$$
The denominators in (\sumrho) tell us that $\rpr$ has operator valued
poles whenever $\g\in\Z$, that is when $\rpr$ acts on a state whose
momentum $k'$ satisfies ${(k\!\cdot\!k')/4}\in\Z$. This means that although
the adjoint action $\dpr\is\ad\rpr$ is a derivation of
$\hbox{\it Vir}\oplus\!\!\hbox{\it Vir}$,
it is not, strictly speaking, an inner derivation of the entire
conformal field theory --- the presence of poles is what makes the
transformations non-trivial. We will make repeated use of the analytic
dependence on the mode shift $\g$.

We already know from [\CNW], as described in the introduction, that
deformations are related to integrals of the $(1,1)$-form
$\phi=\Phi(z,\zbar)dz\!\wedge\!d\zbar$.
Equation (\sumrho) can in fact be written
$$
\rpr=-\oop\Int_{|z|<R}\!d^2z\,\Phi(z,\zbar)
	={1\/2\pi i}\Int_{|z|<R}\!\phi\punkt
	\Eqn\intrho
$$
To establish equality of equations (\sumrho) and (\intrho) we are led
to introduce the fundamental regularization used in this paper.

\section{Regularization}
The regularization we will use is defined through analytic continuation
in the mode shift $\g$. We introduce it by verifying eq\.\ (\intrho).
Explicit calculation of the right hand side yields
$$
\eqalign{-\oop\Int_{|z|<R}\!d^2z\,\Phi(z,\zbar)
	&=-\oop\Int_{|z|<R}\!d^2z\,\sum_{m,n\in\Z}
			\Phi_{mn}|z|^{2\g}z^{-1-m}\zbar^{-1-n}\cr
	&=-\oop\Int_0^R\!rdr\Int_0^{2\pi}\!d\theta\,\sum_{m,n\in\Z}
			\Phi_{mn}r^{2\g-2-m-n}e^{i(n-m)\theta}\cr
	&=-\sum_{n\in\Z}{R^{2(\g-n)}\/\g-n}\Phi_{nn}\punkt\cr}\eqn
$$
The prescription for the radial integration is
$$
\Int_0^1\!dx\,x^\a={1\/1+\a}\komma\quad
	\a\neq-1\komma\Eqn\prescr
$$
and it is obtained through analytic continuation from the true region of
convergence, $\a\!>\!-1$. In terms of the primitive functions,
it corresponds to setting $x^{\a+1}|_{x=0}=0$, while $\log x|_{x=0}$
is undetermined. The only remaining divergences that can not (and should not)
be regularized this way are the logarithmic ones responsible for the
pole in (\prescr).
The prescription for evaluating surface integrals is to first
perform the angular integration (to eliminate potential poles with zero
residue) and then regularize the radial integration according to (\prescr).

Some comments are in order.

This type of regularization is exactly the one used for calculation of
amplitudes in string theory. When calculating e.g\.\ a four-string amplitude
by integrating over the position of one of the vertex operators (this type
of integral will be discussed below), one encounters divergences when it
approaches the locus of one of the others. The actual convergence region
of the integral is a bounded region for the Mandelstam variables, that
does not contain any resonances. Not until the result is analytically
continued in momenta does the pole structure, exhibiting resonances in
the different channels, arise. This analytic continuation amounts
exactly to (\prescr).

Another point worth mentioning is that as a corollary of (\prescr) one has
$$
\Int_0^\infty\!dx\,x^\a=\left(\,\Int_0^1+\Int_1^\infty\,\right)dx\,x^\a
	=\Int_0^1\!dx\,(x^\a+x^{-2-\a})
	={1\/1+\a}+{1\/-1-\a}=0\punkt\Eqn\intinfzero
$$
We should stress here that although the calculation is not valid
for $\a\is\-1$ (there is a delta function interpretation), the formalism
allows us to care about the integral only as an analytic function of $\a$.
Then the singularity at $\a\is\-1$ is removable, and eq\.\ (\intinfzero)
is valid for all $\a$ through analytic continuation.
For a field $\Phi(z,\zbar)$ on the the complex plane with singularities
only at $z\is 0$ and $\infty$, the statement (\intinfzero) translates into
$$
\Int_\C \!d^2z\,\Phi(z,\zbar)=0\punkt\Eqn\intzero
$$

If one is not used to calculating string amplitudes, the prescription
(\prescr) may look far-fetched. One may then consider its meaning in
cylindrical coordinates, where time is Wick-rotated back to Minkowski
signature. Then eq\.\ (\intinfzero) amounts to
$\int_{-\infty}^\infty\!dt\,e^{i\b t}\is 0$ as an analytic function of $\b$,
which is a less surprising statement.

Furthermore, this regularization is closely related to $\zeta$-function
regularization, that is commonplace in string theory, in that both just
define analytic continuations of sums or integrals away from the regions
of convergence. We will comment on this connection later.

Finally, the analytic behaviour of $\rpr$ in $\g$ makes it unnecessary to
keep track of the ``ill-definedness'' of $\rpr$ as an operator on the
string Hilbert space. It is analytic almost everywhere, and its behaviour at
the singular points is well controlled. Every calculation may be performed
{\it as if} $\ad\rpr$ where an inner derivation.

\section{Transformation of Operators}
The form (\intrho) of the generator of a deformation $\rpr$ opens for
calculations of commutators as integrals of correlation functions over the
complex plane, instead of making direct use of the mode expansions,
much in the same spirit as one uses contour integrals in a holomorphic
theory. The transformation of a local field $\Psi(z,\zbar)$ is
$$
\delta_\Phi\Psi(z,\zbar)=\Bigl[\,-\oop
	\Int_{|w|<|z|}\!d^2w\,\Phi(w,\wbar)\,\komma\,\,
	\Psi(z,\zbar)\,\,\Bigr]\punkt\eqn
$$
Using the property (\intzero) this is rewritten as
$$
\eqalign{\delta_\Phi\Psi(z,\zbar)&=\oop\Bigl\{
	\Int_{|w|>|z|}\!d^2w\,\Phi(w,\wbar)\Psi(z,\zbar)
	+\Int_{|w|<|z|}\!d^2w\,\Psi(z,\zbar)\Phi(w,\wbar)\Bigr\}\cr
	&=\oop\Int_\C\!d^2w\,\R\,[\,\Phi(w,\wbar)\Psi(z,\zbar)\,]\komma\cr}
		\Eqn\deltaphipsi
$$
$\R$ denoting radial ordering. This is a desirable expression, since radial
ordering is exactly what is needed in order for the mode expansions of
normal ordering terms to converge. It also means, in the light of (\intzero),
that regular terms in the operator product of (\deltaphipsi) do not
contribute to the commutator.

The drawback of expressions like this, as compared to contour integrals in
a holomorphic theory, is that there is no analogy to deformation of
integration contours. It is not allowed to expand $\Phi(w,\wbar)$ in a
Taylor series around $w\is z$, since such an expansion only converges
inside the circle $|w-z|<|z|$.

The first thing to check is that (\deltaphipsi) gives the correct result
for $\delta_\Phi T(z,\zbar)$ (the calculation for $\bar T$ is analogous).
We thus have
$$
\eqalign{
\delta_\Phi T(z,\zbar)&=\oop\Int_\C\!d^2w\,\R\,[T(z)\Phi(w,\wbar)]\cr
	&=\oop\Int_\C\!d^2w\,\left\{{1\/(z-w)^2}\Phi(w,\wbar)
		+{1\/z-w}\d\Phi(w,\wbar)+\hbox{regular}\right\}\komma\cr}
	\Eqn\deltaphiT
$$
where the explicit functions of $z\!-\!w$ are defined through their
convergent series expansions in the regions $|w|\!<\!|z|$ and $|w|\!>\!|z|$.
Splitting the integration region and expanding the series gives
for each term in the expansion (\modeexp) of $\Phi$ an integral of the type
$$
\eqalign{
J(\a,m;n;z)&=\oop\Int_\C\!d^2w\,|w|^{2\a}w^m(z-w)^n\cr
	&=|z|^{2(\a+1)}z^{m+n}\oop\Int_\C\!d^2w\,|w|^{2\a}w^m(1-w)^n
	=|z|^{2(\a+1)}z^{m+n}J_0(\a,m;n)\punkt\cr}\Eqn\IIO
$$
We calculate $J_0(\a,m;n)$ as
$$
\eqalign{
J_0(\a,m;n)=&\oop\Int_{|w|<1}\!d^2w\,\sum_{k=0}^\infty
		(-1)^k{n\choose k}|w|^{2\a}w^{m+k}\cr
	+&\oop\Int_{|w|>1}\!d^2w\,\sum_{k=0}^\infty
		(-1)^{n+k}{n\choose k}|w|^{2\a}w^{m+n-k}\punkt\cr}\eqn
$$
The first integral contributes only when $m\!\leq\!0$, its value is then
${(-1)^m\/1+\a}{n\choose -m}$; the second one when $m\!+\!n\!\geq\!0$ with the
value $-{(-1)^m\/1+\a}{n\choose n+m}$. It is obvious that the two terms
cancel when $n\!\geq\!0$, i.e\.\ when the integrand only has singularities
at $w\is 0$ or $\infty$. When $n\is -\!N$, $N\!=1,2,\ldots$, the two terms
combine (for any $\a\!\neq\!-1$ and $m$) to give
$$
J_0(\a,m;-N)={1\/1+\a}{(1-m)_{N-1}\/(N-1)!}\Eqn\jnoll
$$
(see Appendix for notation),
which in particular means that if we let $f(z,\zbar)\is |z|^{2\a}z^m$,
we obtain
$$
\oop\Int_\C\!d^2w\,
	\left\{{1\/(1-w)^2}f(w,\wbar)+{1\/1-w}\d f(w,\wbar)\right\}
	={1-m\/1+\a}+{\a+m\/1+\a}=1\komma\eqn
$$
and, in view of (\IIO),
$$
\oop\Int_\C\!d^2w\,
	\left\{{1\/(z-w)^2}f(w,\wbar)+{1\/z-w}\d f(w,\wbar)\right\}
	=|z|^{2(\a+1)}z^{m-2}={\zbar\/z}f(z,\zbar)\komma\eqn
$$
thus verifying equation (\Tvar) for the variation of the stress tensor.

The next natural thing to examine is how physical vertex operators deform.
Equation (\deltaphipsi) contains this information, although in rather implicit
form. Any explicit formula will depend on the detailed behaviour of
the operator product $\Phi(w,\wbar)\Psi(z,\zbar)$. If we also take $\Psi$
to be a physical vertex operator, $\delta_\Phi\Psi(z,\zbar)$ is the
operator containing the four-string amplitudes with any two states besides
$\Phi$ and $\Psi$. Up to a constant:
$$
|z|^{-2}A_4(\V,\V',\Phi,\Psi)=\oop\Int_\C\!d^2w
	\bra\V|\R\,[\Phi(w,\wbar)\Psi(z,\zbar)]|\V'\ket
	=\bra\V|\delta_\Phi\Psi(z,\zbar)|\V'\ket\punkt\Eqn\Afour
$$
Since no Taylor expansion around $w\is z$ is allowed, we have to evaluate
the integral for each mode of $\Phi$. We perform this calculation for
the deformation of a tachyon by another tachyon as an example. Any other
case goes along the same lines, and no technical difficulties are left
out by this example. In order to give a more general formula, one would
have to resort to the DDF construction [\DDF] of physical vertex operators
for states of arbitrary $m^2$.

Let the two tachyon vertices be $\Phi(z,\zbar)\is\exp(ik\!\cdot\!X(z,\zbar))$
and $\Psi(z,\zbar)\is\exp(ik'\!\cdot\!X(z,\zbar))$, with
$k^2\is k'^2\is 8$. The operator product is
$$
\Phi(w,\wbar)\Psi(z,\zbar)=|z-w|^{k\cdot k'\/2}
	e^{ik\cdot X(w,\wbar)+ik'\cdot X(z,\zbar)}\punkt\Eqn\tachOP
$$
The integrals to be evaluated are of the type
$$
\eqalign{
I(\a,m;\b,n;z)&=\oop\Int_\C\!d^2w\,|w|^{2\a}w^m|z-w|^{2\b}(z-w)^n\cr
	&=|z|^{2(1+\a+\b)}z^{m+n}
	\oop\Int_\C\!d^2w\,|w|^{2\a}w^m|1-w|^{2\b}(1-w)^n\cr
	&=|z|^{2(1+\a+\b)}z^{m+n}I_0(\a,m;\b,n)\punkt\cr}
	\Eqn\Inolldef
$$
In our specific example we have $n\is 0$, but a generic operator product
will involve the general form.

The value of $I_0(\a,m;\b,n)$ is known [\Heterotic]. It is
$$
I_0(\a,m;\b,n)={\G(1+\a+m)\G(1+\b+n)\G(-1-\a-\b)\/
		\G(-\a)\G(-\b)\G(2+\a+\b+m+n)}\punkt\Eqn\useful
$$
In ref\.\ [\Heterotic], this integral was calculated by the standard method
for evaluating tree-level string amplitudes.
In principle, $I_0(\a,m;\b,n)$ may also be calculated using series
expansions for the functions of $z\!-\!w$ and splitting the integration
region in $|w|\!<\!|z|$ and $|w|\!>\!|z|$. Then the calculation reads
$I_0\is I_<\!+\!I_>$, and
$$
I_<(\a,m;\b,n)=\oop\Int_{|w|<1}\!d^2w\,|w|^{2\a}w^m
	\sum_{k,l=0}^\infty(-1)^{k+l}{\b+n\choose k}{\b\choose l}w^k\wbar^l
	\punkt\eqn
$$
The angular integration restricts the sum to those terms which have
$m\!+\!k\is l$. One has to distinguish the cases $m\!\geq\!0$ and
$m\!\leq\!0$. The remaining sum is collected in a hypergeometric function
(see Appendix).
The calculation of $I_>$ is analogous, and the complete result is
$$
\eqalign{
&I_0(\a,m;\b,n)=\left\{\matrix{{\G(-\b+m)\/\G(-\b)\G(1+m)}
	\Int_0^1\!dx\,x^{\a+m}\F(\-\b\-n,\-\b\+m;1\+m;x)&(m\geq 0)\hfill\cr
	{\G(-\b-m-n)\/\G(-\b-n)\G(1+m)}
	\Int_0^1\!dx\,x^{\a}\F(\-\b,\-\b\-m\-n;1\-m;x)&(m\leq 0)\hfill\cr}
		\right.\cr
&+\left\{\matrix{(-1)^n{\G(-\b+m)\/\G(-\b-n)\G(1+m+n)}
	\Int_0^1\!dx\,x^{-2-\a-\b}\F(\-\b,\-\b\+m;1\+m\+n;x)
		\hfill\cr\hfill (m\+n\geq 0)\cr
	(-1)^n{\G(-\b-m-n)\/\G(-\b)\G(1-m-n)}
	\Int_0^1\!dx\,x^{-2-\a-\b-m-n}\F(\-\b\-n,\-\b\-m\-n;1\-m\-n;x)
		\cr\hfill (m\+n\leq 0)\cr}
		\right.\cr
&\cr
&=\left\{\matrix{{1\/1+\a+m}\,{\G(-\b+m)\/\G(-\b)\G(1+m)}
		\FF(\-\b\-n,\-\b\+m,1\+\a\+m;1\+m,2\+\a\+m;1)\hfill
			&(m\geq 0)\hfill\cr
		{1\/1+\a}\,{\G(-\b-m-n)\/\G(-\b-n)\G(1-m)}
		\FF(\-\b,\-\b\-m\-n,1\+\a;1\-m,2\+\a;1)\hfill
			&(m\leq 0)\hfill\cr}\right.\cr
&\cr
&+\left\{\matrix{{(-1)^{n+1}\/1+\a+\b}\,{\G(-\b+m)\/\G(-\b-n)\G(1+m+n)}
		\FF(\-\b,\-\b\+m,-1\-\a\-\b;1\+m+n,\-\a\-\b;1)\hfill\cr
			\hfill(m\+n\geq 0)\phantom{\punkt}\cr
		{(-1)^{n+1}\/1+\a+\b+m+n}\,{\G(-\b-m-n)\/\G(-\b)\G(1-m-n)}
		\hfill\cr
		\phantom{XXX}\times\FF(\-\b\-n,\-\b\-m\-n,-1\-\a\-\b\-m\-n;
			1\-m\-n,\-\a\-\b\-m\-n;1)\hfill\cr
			\hfill(m\+n\leq 0)\punkt\cr}\right.\cr}
\Eqn\hyperint
$$
The regularization is now hidden in the transition from hypergeometric
{\it series} to hypergeometric {\it function}, the latter being an analytic
continuation of the former, identical to $\zeta$-function regularization
of the sum.

While the value of $\F$ at $1$ is given by the Gauss formula,
$$
\F(a,b;c;1)={\G(c)\G(c-a-b)\/\G(c-a)\G(c-b)}\komma\eqn
$$
the general value of $\FF$ at $1$ can not be expressed in terms of simpler
functions, such as the $\G$-function, apart from some special cases [\Hyper].
The integrals we have at hand do not seem to belong to these. On the
other hand, an alternative calculation [\Heterotic]
has already shown that the sum
$I_<+I_>$ simplifies to (\useful). The difficulty with the calculation
we are trying to perform is that the pole structure at $w\is 1$ is treated
in an asymmetric fashion. The boundary line between the two integration
regions goes through $w\is 1$, so that each of $I_<$ and $I_>$ aquires
poles from integration over a small semicircle around $w\is 1$. Many
of these poles cancel between the two integrals.
By examining the behaviour of the integrands in (\hyperint) at $w\is 1$,
using the technique of the Appendix,
we see that there are poles for half-integer values of $\b$,
but when the contributions from $I_<$ and $I_>$ are added, only the
integer ones survive. They correspond to the physical resonances
in $\delta_\Phi\Psi$. The residues of the poles can be expressed as
finite sums, and we have checked our calculation by comparing these
with the pole structure of (\useful).

We would like to remark on a curious and important fact.
The integral $J_0(\a,m;n)$
already calculated seems to be identical to $I_0(\a,m;0,n)$, so one
would expect the former to be obtained from the latter as
$\lim_{\b\rightarrow 0}I_0(\a,m;\b,n)$, as an analytic function of $\a$.
For $n\!\geq\!0$ this gives
identically zero, but for $n\is-N$, $N\is 1,2,\ldots$ there is a removable
singularity due to the presence of poles both in the nominator and the
denominator. The unique limit is
$$
\lim_{\b\rightarrow 0}I_0(\a,m;\b,-N)={1\/1+\a}{(-\a-m)_{N-1}\/(N-1)!}
	\komma\eqn
$$
which is obviously not in agreement with (\jnoll). The difference can be
interpreted as follows. The procedure of splitting the integration region
may be viewed as omitting the annulus
$$D_\e=\{w\in\C\,\,|\,\,1\-\e<|w|<1\+\e\}\eqn$$
from the integration region, and letting $\e\!\rightarrow\!0$.
Since $D_\e$ contains the point $w\is 1$ where we have potential
singularities, it is not obvious that the integral over $D_\e$ will
vanish in the limit $\e\!\rightarrow\!0$. We can exemplify this with
the integrand $(1\-w)^{-2}$, where it is not too difficult to evaluate
the integral explicitly in the limit $\e\!\rightarrow\!0$
($D_\e$ can effectively be replaced
by $\{w\,|\,1\-\e<\Re w<1\+\e\}$ and by symmetry this can in turn be reduced
to $\{w\,|\,1\-\e<\Re w<1\+\e\,,\,|\Im w|>\e\}$).
We obtain
$$
\lim_{\e\rightarrow 0}\oop\Int_{D_\e}\!{d^2w\/(1-w)^2}=-1\komma\eqn
$$
which exactly matches the difference between $I_0(0,0;0,-2)$ and
$J_0(0,0;-2)$, thus adding support to our interpretation.
The choice of evaluating $J_0$ as we did, i.e., of effectively
integrating over $\C\,\backslash D_\e$, has a simple physical
motivation --- it corresponds to radial point-splitting of the
operators in the commutator. In the generic analytic case it is seen
to be of no importance, but for the deformation of $T(z)$ it was crucial:
while momenta are good variables for analytic continuation, the conformal
dimension of $T(z)$ is not. We would like to emphasize clearly that
the physically correct result is always produced by splitting the integral
in $I_<$ and $I_>$, as implied by the series expansion of the normal
ordering terms.
Without giving a strict argument, discrepancies
between the analytic expression and the result of a point-split series
expansion should occur exactly when the difference between the inner and
outer expansions has a meaning as a non-vanishing distribution on the
circle $|w|\is 1$.

After this excursion, we return to the example with tachyonic deformation
of a tachyon. We replace the scalar products of momenta by Mandelstam
variables according to
$$
\eqalign{
	&s=-(k+p)^2\komma\cr
	&t=-(k+k')^2\komma\cr
	&u=-(k'+p)^2\komma\cr}
\eqn
$$
and define $p'\is-k\-k'\-p$.
Then $s\+t\+u\+p^2\+p'^2\+16\is 0$
(note that $p$ and thus $s$ and $u$ are operator valued).
Inserting the integral (\useful) in (\deltaphipsi) using
the operator product (\tachOP) gives the result
$$
\eqalign{
\delta_\Phi\Psi(z,\zbar)=&\sum_{m,n}
	{\G(-{s\/8}-1-m-{p^2\/8})\G(-{t\/8}-1)\G(-{u\/8}+1+n-{p'^2\/8})\/
		\G({s\/8}+2+n+{p^2\/8})\G({t\/8}+2)\G({u\/8}-m+{p'^2\/8})}\cr
	&\phantom{XXXXXXXXXXX}\times\,:\Phi_{mn}\Psi(z,\zbar):\,
		|z|^{{u\/4}+{p'^2\/4}}z^{-1-m}\zbar^{-1-n}\cr\cr
	=&\sum_{M,N,m,n}
	{\G(-{s\/8}-1-m-{p^2\/8})\G(-{t\/8}-1)\G(-{u\/8}+1+n-{p'^2\/8})\/
		\G({s\/8}+2+n+{p^2\/8})\G({t\/8}+2)\G({u\/8}-m+{p'^2\/8})}\cr
	&\phantom{XXXXXXXXXXX}\times\,:\Phi_{mn}\Psi_{M-m,N-n}:\,
		|z|^{-4-{p^2\/4}+{p'^2\/4}}z^{-1-M}\zbar^{-1-N}\punkt\cr}
\Eqn\tachtach
$$
This operator contains as its $|z|^{-2}$-component all four-string amplitudes
of two tachyons and any other two physical states. This term has
$M\is N\is-\Delta\-2$, where $\Delta\is{1\/8}(p^2\-p'^2)$ is the shift in
excitation number from the in-state to the out-state. It is easy to show,
using $\G(x)\G(1-x)\is\pi/\sin(\pi x)$, that it is symmetric under the exchange
of $\Phi$ and $\Psi$, i.e., the quotient of $\G$-functions in eq\.\ (\tachtach)
is invariant under $m\!\leftrightarrow\!-\Delta\-2\-m$,
$n\!\leftrightarrow\!-\Delta\-2\-n$, $s\!\leftrightarrow\!u$.
Especially it contains the well-known four-tachyon amplitude at
$M\is N\is-2$, $m\is n\is-1$:
$$
A_{\hbox{\eightrm 4-tachyon}}(s,t,u)
	\sim{\G(-{s\/8}-1)\G(-{t\/8}-1)\G(-{u\/8}-1)\/
	\G({s\/8}+2)\G({t\/8}+2)\G({u\/8}+2)}\punkt\eqn
$$

A more general class of operators that naturally enter string amplitudes
are the multi-local operators of the type
$$
\W^{(N)}(z_1,\zbar_1;\ldots;z_N,\zbar_N)=
	\R\,[\,\,\prod_{i=1}^N\V_i(z_i,\zbar_i)\,]\punkt
\Eqn\wdef
$$
Consider the product (\wdef) with $|z_1|>|z_2|>\ldots>|z_N|$. Then, using
(\deltaphipsi) and (\intzero), and letting each local field transform as
$$
\eqalign{
	&\delta_\Phi\W^{(N)}(z_1,\zbar_1;\ldots;z_N,\zbar_N)
	=\sum_{i=1}^N\V_1(z_1,\zbar_1)\ldots\delta_\Phi(|z_i|)
		\V_i(z_i,\zbar_i)\ldots\V_N(z_N,\zbar_N)\cr
	&\cr
	&=\oop\Int_{|w|>|z_1|}\!d^2w\,
		\Phi(w,\wbar)\V_1(z_1,\zbar_1)\ldots\V_N(z_N,\zbar_N)\cr
	&+\oop\sum_{i=1}^{N-1}
		\V_1(z_1,\zbar_1)\!\ldots\!\V_i(z_i,\zbar_i)
		\Bigl(\!\!\Int_{|w|<|z_i|}
		\!\!+\!\!\Int_{|w|>|z_{i+1}|}\!\!\!\Bigr)d^2w\,\Phi(w,\wbar)
		\V_{i+1}(z_{i+1},\zbar_{i+1})\!\ldots\!\V_N(z_N,\zbar_N)\cr
	&+\oop\Int_{|w|<|z_N|}\!d^2w\,
		\V_1(z_1,\zbar_1)\ldots\V_N(z_N,\zbar_N)\Phi(w,\wbar)\cr
	&\cr
	&=\oop\Int_{|w|>|z_1|}\!d^2w\,
		\Phi(w,\wbar)\V_1(z_1,\zbar_1)\ldots\V_N(z_N,\zbar_N)\cr
	&+\oop\sum_{i=1}^{N-1}
		\Int_{|z_{i+1}|<|w|<|z_i|}d^2w
		\V_1(z_1,\zbar_1)\ldots\V_i(z_i,\zbar_i)\Phi(w,\wbar)
		\V_{i+1}(z_{i+1},\zbar_{i+1})\ldots\V_N(z_N,\zbar_N)\cr
	&+\oop\Int_{|w|<|z_N|}\!d^2w\,
		\V_1(z_1,\zbar_1)\ldots\V_N(z_N,\zbar_N)\Phi(w,\wbar)\cr
	&\cr
	&=\oop\Int_\C\!d^2w\,\R\,[\,\Phi(w,\wbar)
		\prod_{i=1}^N\V_i(z_i,\zbar_i)\,]
		\punkt\cr}\Eqn\multidef
$$
This means that the $N$-string amplitudes pick up variations containing
the $(N\+1)$-string amplitudes, under the condition that the vacua at $R\is 0$
and $R\is\infty$ do not transform (which actually follows from our
regularization). It is also obvious that an argument for background invariance
[\Background]
has to involve transformations on the multi-string Fock space
--- no transformation of the external states at $R\is 0,\infty$ contracting
(\wdef) can compensate for the transformations (\multidef).
As amplitudes are constructed from operators at different radii, the
possibility of treating deformations as ``almost inner automorphisms'' is
not relevant for invariance of expectation values of expressions like
(\wdef) unless all operators and states reside at equal radii. Equation
(\multidef) should provide the natural connection from our canonical
first-quantized formalism to transformations on the string field theory
Fock space.

It is important to note that eq\.\ (\multidef) relies directly on
the regularization used.
It is of course the most natural thing to write down, but in other
regularization schemes it will not be consistent with the transformations
of local fields.

\section{Commutators of Deformations}
We have treated the deformations as though they were inner derivations
of the conformal field theory. However, the resonances occurring for
deformations of physical vertex operators imply that this is not really
true --- the resonances are exactly what makes the transformations
non-trivial, and the deformed theory inequivalent to the undeformed
one, simply in the sense that they are not related by an automorphism.
The deformations as we have defined them,
constitute the unique prescription (modulo inner derivations)
for parallel transport of states and
operators that respect conformal invariance. The resonances are
essential, and necessary to make the deformations non-trivial.

The corresponding connection has several interesting local
properties. An important consequence of
our regularization prescription, namely of the result (\intinfzero), is that
the
connection is compatible with the Zamolodchikov metric:
$$
	\delta_\Phi{\bf 1}=\oop\Int_\C\!d^{2}z\Phi(z,\zbar)=0\punkt
\Eqn\metriccomp
$$
The curvature of the connection is contained in the commutators of any two
deformations. The necessary calculations have been performed in
the previous chapter and we find that the curvature vanishes:
$$
\eqalign{
	&[\,\delta_\Phi,\delta_\Psi\,]\W^{(N)}(z_1,\zbar_1;\ldots;z_N,\zbar_N)=
\cr
	&=\delta_\Phi\oop\Int_\C\!d^2w\,\R\,[\,\Psi(w,\wbar)
		\prod_{i=1}^N\V_i(z_i,\zbar_i)\,]-
	\delta_\Psi\oop\Int_\C\!d^2w\,\R\,[\,\Phi(w,\wbar)
		\prod_{i=1}^N\V_i(z_i,\zbar_i)\,]\cr
	&=\oopsq\Int_\C\!d^2w\Int_\C\!d^2z\,\R\,[\,(\Phi(w,\wbar)\Psi(z,\zbar)
	-\Psi(w,\wbar)\Phi(z,\zbar))
		\prod_{i=1}^N\V_i(z_i,\zbar_i)\,]=0\punkt}
\Eqn\phipsi
$$

It is important to note that if we
would have computed the commutator between two deformations
in a strictly first-quantized
formalism then we would have been led to the commutator between
the two generators of the form (\intrho). This commutator is
however non-zero:
$$
\eqalign{
	[\,\rpr,\rqr\,]&={1\/\pi^2}\Int_{\ss|z|<R\atop\ss|w|<R}\!d^2zd^2w\,
		[\,\Phi(w,\wbar),\Psi(z,\zbar)\,]\cr
	&={1\/\pi^2}\Int_{\ss|z|>R\atop\ss|w|<R}\!d^2zd^2w\,
		\Psi(z,\zbar)\Phi(w,\wbar)
		-{1\/\pi^2}\Int_{\ss|z|<R\atop\ss|w|>R}\!d^2zd^2w\,
		\Phi(w,\wbar)\Psi(z,\zbar)\cr
	&={1\/\pi^2}\Bigl(\Int_{\ss|z|>R\atop\ss|w|<R}-
			\Int_{\ss|z|<R\atop\ss|w|>R}\Bigr)\,d^2zd^2w\,
		\R\,[\,\Phi(w,\wbar)\Psi(z,\zbar)\,]\cr
	&={1\/\pi^2}\Bigl(\Int_{\ss|z|\in\C\atop\ss|w|<R}-
			\Int_{\ss|z|<R\atop\ss|w|\in\C}\Bigr)\,d^2zd^2w\,
		\R\,[\,\Phi(w,\wbar)\Psi(z,\zbar)\,]\cr
	&=\r_{\delta_\Phi\Psi-\delta_\Psi\Phi}(R)\punkt\cr}
\Eqn\rhorho
$$
{}From the fact that the amplitude (\Afour) is independent of $z$ for physical
$|\V\ket$, $|\V'\ket$, it follows that all $z$ dependent components of
$\delta_\Phi\Psi(z,\zbar)$ are trivial operators, i.e., vanish on the
physical Hilbert space (this space being considered as consisting of
equivalence classes of physical states modulo trivial states).
Using the symmetry of physical amplitudes, one concludes that the entire
operator $\delta_\Phi\Psi(z,\zbar)-\delta_\Psi\Phi(z,\zbar)$ is trivial
but nevertheless non-zero.

The curvature tensor (\phipsi) thus vanishes on the full state space, which
of course means that its pullback to the physical state space vanishes.
If one instead considers the pullback of the connection
$\delta_\Phi$ to the physical state space, one finds that it carries
non-zero curvature, which is consistent with what is known concerning
the Zamolodchikov metric.

Further examination in the second-quantized formalism
of the local properties of the moduli space of
conformal field theories, such as the
finite parallel transport generated by the
connection and the construction of a connection for the
tangent bundle of the moduli space, will be deferred to [\vongussichsundell].

\vskip6\parskip
\noindent{\tencp Acknowledgement:} M.C. would like to thank
Mark Evans and M\aa ns Henningson for generously sharing opinions
on the subject during the initial stage of this work.
\vskip2\parskip
\vfill\eject
\appendix{Hypergeometric Functions}
The hypergeometric series ${}_AF_B(a_1,\ldots,a_A;b_1\ldots,b_B;z)$ [\Hyper]
is defined as
$$
\eqalign{
{}_AF_B(a_1,\ldots,a_A;b_1\ldots,b_B;z)&=
	\sum_{n=0}^\infty{(a_1)_n\ldots(a_A)_n\/
		n!\,(b_1)_n\ldots(b_B)_n}\,z^n\cr
	&={\G(b_1)\ldots\G(b_B)\/\G(a_1)\ldots\G(a_A)}
	\sum_{n=0}^\infty{\G(a_1+n)\ldots\G(a_A+n)\/
		n!\,\G(b_1+n)\ldots\G(b_B+n)}\,z^n\komma\cr}
\eqno{(\A.1)}
$$
where $(x)_k$ is the Pochhammer symbol
$(x)_k=x(x+1)\ldots(x+k-1)$. The relevant hypergeometric functions in
the present application are those with $A\is B\+1$. The hypergeometric
functions constitute the analytic continuation of the hypergeometric
series (A.1) outside of the convergence region of the series.
They are analytic for $z\!\in\!\C$, except for the possibility of
poles at $z\is 0$ and a cut from
$z\is 1$ to $z\is\infty$ (depending on the values of the parameters).
The asymptotic properties of the series (A.1) can be investigated using
the asymptotic behaviour of the $\G$-function,
$$
\log\G(n)=(n-{1\/2})\log n-n+\log\sqrt{2\pi}
	+\sum_{k=1}^\infty{B_{2k}\/2k(2k-1)}n^{1-2k}\komma
\eqno{(\A.2)}
$$
where $B_{2k}$ are Bernoulli numbers.
This asymptotic expansion applied to the quotient of two $\G$-functions
gives an expansion
$$
{\G(n+a)\/\G(n+b)}=n^{a-b}\sum_{k=0}^\infty c_k(a,b)n^{-k}\punkt
\eqno{(\A.3)}
$$
The class of hypergeometric series having interesting convergence
properties is therefore the case $A\is B\+1$. The convergence
criterion at $z\is 1$ is
$\raise.8pt\hbox{$\sum$} a-\raise.8pt\hbox{$\sum$} b<0$.
The analytic continuation
can be seen as $\zeta$-function regularization of the series. The
pole of the $\zeta$-function
$\zeta(s)=\raise.8pt\hbox{$\sum_{n=1}^\infty$}n^{-s}$ at
$s\is 1$ with residue $1$ gives poles at
$\raise.8pt\hbox{$\sum$} a-\raise.8pt\hbox{$\sum$}\,b=0,1,2\ldots$
for the hypergeometric funtion
at $z\is 1$, i.e., exactly when
the sum $\raise.8pt\hbox{$\sum$} {1\over n}$ occurs from (A.3).

The value of $\F$ at $z\is 1$ is given by the Gauss formula
$$
\F(a,b;c;1)={\G(c)\G(c-a-b)\/\G(c-a)\G(c-b)}\komma\eqno{(\A.4)}
$$
but the corresponding general value of ${}_{A+1}F_A$ at $z\is 1$ can not be
expressed in terms of simpler functions for generic values of the
$a$ and $b$ parameters. During the evaluation of some integrals in section 4,
we wanted to find the value of
$$
\Int_0^1\!dx\,x^p\F(a,b;c;x)={1\/p+1}\FF(a,b,p+1;c,p+2;1)\punkt\eqno{(\A.5)}
$$
We can derive an expression for the pole structure of (A.5) as follows.
The lower integration limit is simple, one picks up poles (using the
regularization (\prescr)) when $p\is -N$, $N\is 1,2,\ldots$, with the residues
${(a)_N(b)_N\/N!(c)_N}$. For the upper limit, we may use a ``reflection
formula'' for the hypergeometric function. With the same argument as
above for the value of the function at $z\is 1$, we conclude that
poles only occur when $c\-a\-b\is -N$, $N\is 1,2,\ldots$, and [\Hyper]
$$
\eqalign{
\F(&a,b;a+b-N;x)={\G(N)\G(a+b-N)\/\G(a)\G(b)}\sum_{k=0}^{N-1}
	{(a-N)_k(b-N)_k\/k!\,(1-N)_k}(1-x)^{k-N}\cr
	&-(-1)^N{\G(a+b-N)\/\G(a-N)\G(b-N)}\cr
	&\phantom{XX}\times\sum_{k=0}^\infty
	\bigl[\log(1-x)-\psi(1+k)-\psi(1+N+k)+\psi(a+k)+\psi(b+k)\bigr]
	(1-x)^k\komma\cr}
\eqno{(\A.6)}
$$
where $\psi(z)$ is the digamma function $\psi(z)={d\/dz}\log\G(z)$
(the cut from $z\is 1$ to $z\is\infty$ of the hypergeometric function
becomes logarithmic for these values of the parameters).
The only part of (A.6) that can develop poles upon integration weighted
with a function that is regular at $x\is 1$ is the finite sum. We expand
the function $x^p$ in a power series around $x\is 1$ and identify the
coefficient of the $(1-x)^{-1}$-term, that gives the residue of the pole.
The residue is, after a short calculation,
$$
\Res_{c=a+b-N}\Int_0^1\!dx\,x^p\F(a,b;c;x)=
	{\G(a+b-N)(-p)_{N-1}\/\G(a)\G(b)}\sum_{k=0}^{N-1}
	{(a-N)_k(b-N)_k\/k!\,(2+p-N)_k}\punkt\eqno{(\A.7)}
$$
In terms of the integrals $I_<$ and $I_>$ of section 4, we have as an example
$$
\Res_{\b=-{1+N+n\/2}}I_<(\a,m\geq 0;\b,n)=
	{(-\a-m)_{N-1}\/2\G({1+N+n\/2})\G({1+N-n\/2})}
	\sum_{k=0}^{N-1}{\left({1-N-n\/2}\right)_k\left({1-N+n\/2}+m\right)_k
		\/k!\,(2+\a+m-N)_k}\punkt
\eqno{(\A.8)}
$$
We note that the factors in front of the sum makes the residue vanish for
$N\+n$ odd, $|n|\!>\!N$. In fact, the two terms in equation (\hyperint) cancel
for $N\+n$ even, so the remaining poles lie at
$n\is \-N\+1,\-N\+3,\ldots,N\-3,N\-1$, i.e., only for negative integer $\b$.
We are convinced that the finite sums
for the residues can be simplified, but since the result for the integral
already is known, we have contended ourselves to check the equality of (A.8)
with the residues of the known formula (\useful) using {\it Mathematica}.
We remind that although (\useful) gives the correct analytic expression
for the integral, the limit of this analytic function
does not give the right result in some
(a discrete set of) cases. This is due to radial point-splitting of
operator products, as introduced in section 4.
This discrepancy occurred for the deformation of the stress tensor,
whose conformal
dimension is not a good variable for analytic continuation.
The equality of equations (\hyperint) and (\useful) does not seem to
be known in the mathematics literature.
\vfill\eject

\xrm\refout\rm

\end